\documentclass[aps]{revtex4}

\usepackage{graphicx}

\begin{document}

\title{Demonstration of superluminal effects in an absorptionless,
non-reflective system}
\author{D. R. Solli}
\affiliation{Department of Physics, University of California, Berkeley, CA 94720-7300.}
\author{C. F. McCormick}
\affiliation{Department of Physics, University of California, Berkeley, CA 94720-7300.}
\author{C. Ropers}
\author{J. J. Morehead}
\author{R. Y. Chiao}
\affiliation{Department of Physics, University of California, Berkeley, CA 94720-7300.}
\author{J. M. Hickmann}
\email{hickmann@loqnl.ufal.br}
\affiliation{Department of Physics, University of California, Berkeley, CA 94720-7300.}
\affiliation{Departamento de F\'{\i}sica, Universidade Federal de Alagoas, CidadeUniversit%
\'{a}ria, 57072-970, Macei\'{o}, AL, Brazil.}

\begin{abstract}
We present an experimental and theoretical study of a simple, passive system
consisting of a birefringent, two-dimensional photonic crystal and a
polarizer in series, and show that superluminal dispersive effects can arise
even though no incident radiation is absorbed or reflected. We demonstrate
that a vector formulation of the Kramers-Kronig dispersion relations
facilitates an understanding of these counter-intuitive effects.
\end{abstract}

\maketitle

Superluminal group velocities have been observed in a number of different
physical systems. These include passive absorptive \cite{Chu1982}, passive
reflective \cite{Steinberg1993,Solli2003A}, and active transparent \cite%
{Wang2000} media. There have also been numerous theoretical and experimental
proposals to observe superluminality in the tunneling of electromagnetic
wavepackets \cite{Nimtz1997}.

Here we report the first experimental observation of superluminal effects in
a passive system with neither absorption nor reflection. The effects arise
because of a transfer of energy or interference between two modes of the
electromagnetic field, in this case two different polarizations of light. We
are able to interpret these results using a new vector formulation of the
Kramers-Kronig relations.

It is widely believed that the Kramers-Kronig (K-K) relations \cite{Kronig1926}
require that passive systems be either absorptive or reflective in order to exhibit
superluminal effects. In this paper, we demonstrate that this is not the
case; while absorptive and reflective systems are required to have spectral
regions of anomalous dispersion, the converse is not necessarily true. In
fact, the exchange of energy between modes is a \emph{sufficient} condition
for superluminal propagation in any system. We show that these effects are
consistent with causality.

Our experimental system consists of a slab of highly birefringent
two-dimensional (2D) photonic crystal and a linear polarizer, placed in
series. The photonic crystal has fundamental and second-order photonic band
gaps in the regions of 10 and 20 GHz, and displays strong birefringence with
very high transmission in the frequency range between the two gaps \cite%
{Solli2003C}. The crystal itself is an 18-layer hexagonal array of hollow
acrylic rods (outer diameter $1/2"$) with an air-filling fraction (AFF) of
0.60. The crystal was constructed using a method which we have previously
described \cite{Hickmann2002}.

We studied the transmission and dispersive properties of this system between
the two band gaps, using an HP 8720A vector network analyzer (VNA).
Microwaves were coupled to and from free space with polarization-sensitive
horn antennae. The photonic crystal was placed in the far-field of the
transmitter horn to ensure that planar wavefronts of a well-defined
polarization were incident on it. The receiver horn was positioned
immediately behind the photonic crystal on a direct line of sight with the
transmitter horn. In addition, the crystal and receiver horn were placed
inside a microwave-shielded box with an open square aperture, whose size was
chosen to minimize diffraction effects while eliminating signal leakage
around the crystal. This method has proven very effective for transmission
measurements at centimeter wavelengths \cite{Hickmann2002}.

\begin{figure}
\includegraphics{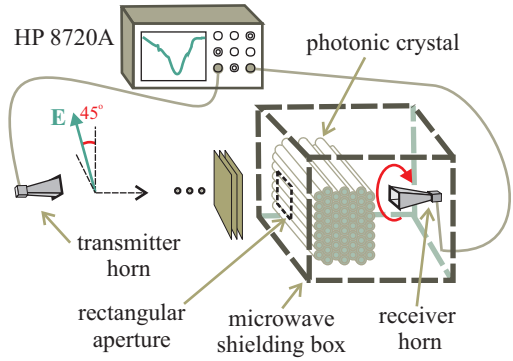}
\caption{Experimental setup.}
\label{newsetup}
\end{figure}

In order to control the polarizations of the incident and detected fields
relative to the fast axis of the crystal, we mounted the transmitter and
receiver horns on precision rotation stages. The angle of the incident
polarization $\theta$ was held fixed at $45^{\circ}$ relative to the fast
axis of the crystal, while the angle of the receiver horn $\beta$ was
allowed to vary. We define our coordinates such that the slab is oriented
with its fast axis parallel to the vertical direction, and label the
incident polarizations TM (transverse magnetic) and TE (transverse electric)
for polarizations parallel and perpendicular to the fast axis, respectively.
Our experimental setup is displayed in Fig. \ref{newsetup}.

\begin{figure}
\includegraphics{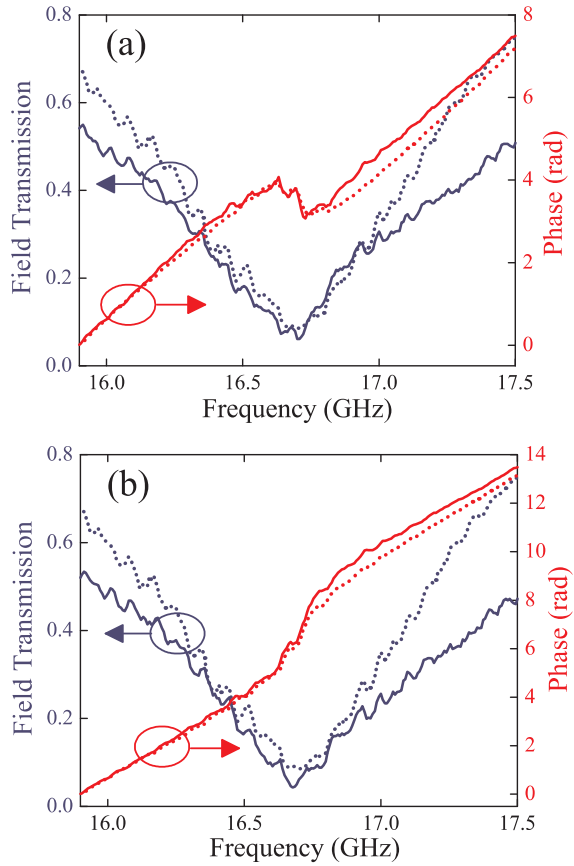}
\caption{Calculated (dotted lines) and measured (solid lines) transmission
(left axis) and phase (right axis) for the detected polarization inclined at 
(a) $40{{}^\circ}$ and (b) $50{{}^\circ}$.}
\label{beta}
\end{figure}

Our transmission amplitude and phase measurements with the receiver horn set
at $40^{\circ }$ and $50^{\circ }$ are shown in Figs. \ref{beta}(a) and \ref%
{beta}(b), respectively. Since this spectral region is far from the band
gaps, the transmission dip is not due to any band gap effect. It is caused
by the fact that the photonic crystal rotates the polarization of the light
by adding a different frequency-dependent phase to each polarization
component \cite{Solli2003C}. For $\beta \leq 45^{\circ }$ there is clear
anomalous dispersion in the vicinity of 16.5-17 GHz (the half-waveplate
frequency for the photonic crystal) while at $\beta \geq 45^{\circ }$ the
dispersion is normal.

A remarkable feature of our data is that while the transmission is identical
for both $\beta =40^{\circ }$ and $\beta =50^{\circ }$, aside from
experimental errors, the phase properties are quite different. Motivated by
these peculiar results, we would first like to derive a simple physical
model capable of explaining these phenomena and illustrating their
connection with causality.

A linear system which is invariant under time translation and is described
by a scalar response (Green's) function $g(t)$ produces a time-dependent
response $b(t)$ to an input $a(t)$ given by the convolution $b(t) =
\int_{\infty}^{-\infty} g(t-\tau )a(\tau )d\tau$. This expression is the
starting point for the usual derivation of the Kramers-Kronig relations \cite%
{Nussenzveig1972}. Our system, however, accepts a time-dependent vector
input and produces a vector output. We must therefore replace the scalar
convolution with the expression

\begin{equation}
b_{i}(t)= \int_{-\infty}^{\infty}g_{ij}(t-\tau)a_{j}(\tau) d\tau =
\int_{-\infty}^{\infty}g_{ij}(t^{\prime})a_{j}(t-t^{\prime})dt^{\prime}
\label{conv}
\end{equation}

\noindent where $b_{i}(t)$ and $a_{i}(t)$ are the $i^{th}$ components of the
time-dependent output and input vectors $\mathbf{b}(t)$ and $\mathbf{a}(t)$, 
$g_{ij}(t)$ is the $(i,j)^{th}$ component of the Green's function matrix
which describes the system. We have employed the standard summation
convention over repeated indices. In the frequency domain, Eq. \ref{conv}
takes the simple form

\begin{equation}
\tilde{B}_{i}(\omega)=\tilde{G}_{ij}(\omega)\tilde{A}_{j}(\omega)
\label{trans}
\end{equation}

\noindent where $\tilde{B}_{i}(\omega)$, $\tilde{A}_{j}(\omega)$, and $%
\tilde{G}_{ij}(\omega)$ are the Fourier transforms of $b_{i}(t)$, $a_{i}(t)$%
, and $g_{ij}(t)$, respectively (assuming these transforms exist).

Causality tells us that the output must vanish for times before the input
has propagated through the system. Thus, we require $\int_{-%
\infty}^{T}g_{ij}(\tau)a_{j}(t-\tau)d\tau = 0$ where $T$ must be greater
than or equal to the relativistic propagation time. Since this relation must
be satisfied for any choice of input, the components of the Green's function
matrix vanish individually for all time prior to $T$. This directly implies
that the components $\tilde{G}_{ij}(\omega)e^{-i\omega T}$ have analytic
continuations for complex frequencies $\tilde{\omega}$ where $\text{Im}(%
\tilde{\omega})>0$ (i.e., the upper half-plane in complex frequency space).
Therefore, it is possible to proceed in the usual way to show that the real
and imaginary parts of each component $\tilde{G}_{ij}(\omega)$ satisfy the
Kramers-Kronig dispersion relations assuming the components $\tilde{G}%
_{ij}(\omega )$ are square integrable:

\begin{equation}
\text{Re} \ \tilde{G}_{ij}(\omega )=\frac{2} {\pi}P\int_{0}^{\infty} \frac{%
\Omega \text{Im} \ \tilde{G}_{ij}(\Omega )} {\Omega ^{2}-\omega ^{2}}d\Omega
\label{kkre}
\end{equation}

\begin{equation}
\text{Im} \ \tilde{G}_{ij}(\omega )=-\frac{2\omega }{\pi }P\int_{0}^{\infty} 
\frac{\text{Re} \ \tilde{G}_{ij}(\Omega )}{\Omega ^{2}-\omega ^{2}}d\Omega
\label{kkim}
\end{equation}

\noindent where $P$ denotes Cauchy's principal value \cite{Nussenzveig1972}.

The signals in our experiment can be expressed as $\mathbf{\tilde{A}}(\omega
)=\tilde{A}(\omega )\mathbf{e}_{A}$ and $\mathbf{\tilde{B}}(\omega )=\tilde{B%
}(\omega )\mathbf{e}_{B}$, where $\mathbf{e}_{A}$ and $\mathbf{e}_{B}$ are
unit vectors in real space, and $\tilde{A}(\omega )$ and $\tilde{B}(\omega )$
are the complex amplitudes of the incident and detected electric fields. In
this special case where $\mathbf{e}_{A}$ and $\mathbf{e}_{B}$ are
frequency-independent,  Eq. \ref{trans} implies that a scalar transfer
function describes the relationship between the complex functions $\tilde{B}%
(\omega )$ and $\tilde{A}(\omega )$

\begin{equation}
\tilde{H}(\omega)=\frac{\tilde{B}(\omega)}{\tilde{A}(\omega)} =
G_{ij}e_{A}^{j}e_{B}^{i}  \label{scalar-t}
\end{equation}

\noindent where $e_{A}^{i}$ and $e_{B}^{i}$ are the $i^{th}$ components of
the unit vectors $\mathbf{e}_{A}$ and $\mathbf{e}_{B}$, respectively. Since
the elements $\tilde{G}_{ij}(\omega)$ satisfy the Kramers-Kronig relations,
it follows that the scalar function $\tilde{H}(\omega)$ must also satisfy
them. However, the components of $\tilde{G}_{ij}(\omega)$ interfere with
each other in $\tilde{H}(\omega)$.

It is this interference which leads to the superluminal effects in our
experiment. The Green's function matrix for our (birefringent) system is

\begin{equation}
\mathbf{\tilde{G}}(\omega )\leftrightarrow \left( 
\begin{array}{cc}
e^{i\phi _{TE}(\omega )} & 0 \\ 
0 & e^{i\phi _{TM}(\omega )}%
\end{array}
\right)  \label{matrix-G}
\end{equation}

\noindent where $\phi _{TE}(\omega)=n_{TE}(\omega) \omega d/c$ and $%
\phi_{TM}(\omega)=n_{TM}(\omega) \omega d/c$ are the frequency-dependent
phases imparted to TE and TM polarizations, $n_{TE}(\omega)$ and $%
n_{TM}(\omega)$ represent the frequency-dependent indices of refraction for
the two polarizations, $d$ is the slab thickness, and $c$ is the vacuum
speed of light. Substituting these matrix elements into Eq. \ref{scalar-t},
we find

\begin{equation}
\tilde{H}(\omega )=\sin (\beta )\sin (\theta )e^{i\phi _{TE}(\omega )}+\cos
(\beta )\cos (\theta )e^{i\phi _{TM}(\omega)}\text{.}  \label{scalar-t3}
\end{equation}

For incident linear polarization at an angle of $\theta =\pi /4$ from the
vertical, the magnitude and phase of the transfer function are, respectively,

\begin{equation}
\left| \tilde{H}(\omega )\right| =\frac{1}{\sqrt{2}}\left\{ 1+\sin (2\beta
)\cos \left[ \Delta \phi (\omega )\right] \right\} ^{1/2}  \label{mag-H}
\end{equation}

\begin{equation}
\arg [\tilde{H}(\omega )]=\arg \left[ e^{i\phi _{TM}(\omega )}\right]
+\arctan \left\{ \frac{\sin [\Delta \phi (\omega )]}{\cos [\Delta \phi
(\omega )]+\cot (\beta )}\right\}  \label{arg-H}
\end{equation}%
where $\Delta \phi (\omega )=\phi _{TE}(\omega )-\phi _{TM}(\omega )$.

The transmission $\left| \tilde{H}(\omega )\right| $ has a minimum for $%
\Delta \phi (\omega _{m})=(2m+1)\pi $ where $\omega _{m}$ is a
half-waveplate frequency of the photonic crystal and $m$ is an integer. For $%
\beta =\pi /4$, $\left| \tilde{H}(\omega _{m})\right| =0$ since the
polarization which emerges from the slab has zero projection along the unit
vector $\mathbf{e}_{B}$. At this point, the phase $\arg \left[ \tilde{H}%
(\omega )\right] $ is undefined since the transmission vanishes. For $\beta
=\pi /4\pm \epsilon $ where $0<\epsilon \ll 1$, we find

\begin{equation}
\left\vert \tilde{H}(\omega _{m})\right\vert =\epsilon +\vartheta (\epsilon
^{3});  \label{mag-H2}
\end{equation}%
\begin{equation}
\left. \frac{\partial \arg [\tilde{H}(\omega )]}{\partial \omega }%
\right\vert _{\omega _{m}}=\pm \frac{1}{2\epsilon }\left. \frac{\partial
\Delta \phi (\omega )}{\partial \omega }\right\vert _{\omega _{m}}+\vartheta
(1)\text{,}  \label{arg-H2}
\end{equation}%
\newline

\noindent where $\tau _{g}=\partial \arg \left[ \tilde{H}(\omega )\right]
/\partial \omega $ is the group delay of the transmitted wave. As $\epsilon
\rightarrow 0$ (i.e., as we approach the singularity in the transfer
function), the transmission goes to zero at $\omega _{m}$ and the group
delay is unbounded. For example, if $\beta =\pi /4-\epsilon $ and $\left.
\partial \Delta \phi (\omega )/\partial \omega \right| _{\omega _{m}}>0$, it
is clear that the group delay can become superluminal, zero, or even
negative depending on the value of $\epsilon $. Thus, if one were to measure
the time of flight of an analytic pulse based on the arrival of its peak,
superluminal results can be obtained. However, there is no violation of
causality here: the group velocity of the pulse has nothing to do with the
signal velocity (i.e., the velocity of \ ``information''), the quantity
restricted by relativity. As we will demonstrate empirically, these
superluminal group velocities are in fact required by causality through the
Kramers-Kronig relations.

In Figs. \ref{beta}(a) and \ref{beta}(b), we have plotted the results of
this simple model described by Eqs. \ref{mag-H} and \ref{arg-H} for
comparison with the experimental results. The frequency-dependent phases $%
\phi _{TE}(\omega )$ and $\phi _{TM}(\omega )$ were calculated using an
indepedent measurement of the indices of refraction of the photonic crystal,
and include phase delays associated with free-propagation in the air-spaces
between the horns and the crystal. It is clear that the model correctly
predicts the behavior of this system. Some discrepancies between the model
and the experiment are observed at the edges of the plotted regions due to
the effects of the band structure of the photonic crystal, which were not
included in the model.

In order to transform between the real and the imaginary parts of $\tilde{H}%
(\omega )$, slightly modified versions of the Kramers-Kronig integrals must
be used since the function $\left| \tilde{H}(\omega )\right| $ is not square
integrable over all real frequencies. In practice, however, one only has
actual data for $\tilde{H}(\omega )$ over a finite range of positive
frequencies between values we label as $\omega _{1}$ and $\omega _{2}$. In
this case, it is possible to apply approximate transforms by truncating the
integrals of Eqs. \ref{kkre} and \ref{kkim} at $\omega _{1}$ and $\omega
_{2} $. These transformations are approximately correct for $\tilde{H}%
(\omega )$ given in Eq. \ref{scalar-t3} if $\omega _{1}\ll \omega \ll \omega
_{2}$.

On the other hand, it seems counter-intuitive, from a Kramers-Kronig point
of view, that the function $\left| \tilde{H}(\omega )\right| $ is symmetric
around $\beta =\pi /4$, while $\arg \tilde{H}(\omega )$ \emph{lacks symmetry}
about this point. In particular, it is usually possible to obtain the phase
of a causal physical response function given its amplitude using the
equation \cite{Toll1956} 
\begin{equation}
\arg \tilde{H}(\omega )=d_{0}\omega -\frac{2\omega }{\pi }\int_{0}^{\infty }%
\frac{\ln \left| \tilde{H}(\Omega )\right| }{\Omega ^{2}-\omega ^{2}}d\Omega
\label{kk-ph}
\end{equation}%
where $d_{0}$ is a fixed, undetermined constant which characterizes the
system in spectral regions of constant transmission. Clearly, this
transformation cannot produce two different phase functions given a single
magnitude. However, it is known that this expression must be modified if the
function $\tilde{H}(\tilde{\omega})$ has zeros in the upper half-plane \cite%
{Toll1956}. Thus, the phase of the response function is not uniquely
determined from a measurement of the amplitude alone; if the response
function has zeros in the upper-half plane, these zeros increase the group
delay in a nontrivial way over what is expected simply from $\left| \tilde{H}%
(\omega )\right| $. For the transfer function given in Eq. \ref{scalar-t3},
we find that the zeros satisfy%
\begin{equation}
\Delta \phi (\tilde{\omega}_{n})=-i\ln \left| \cot (\beta )\right| +2\pi (n+%
\frac{1}{2})  \label{zeros}
\end{equation}%
where $n$ is an integer and $0<\beta <\pi /2$. Given our form of $\Delta
\phi (\omega )$, it follows from Eq. \ref{zeros} that all the relevant zeros
lie in the lower-half plane for $0<\beta <\pi /4$, while all are in the
upper-half plane for $\pi /4<\beta <\pi /2$. Under these conditions, we see
that it should be possible to apply a relatively simple Kramers-Kronig
transformation to obtain the phase of the response function given its
magnitude if $0<\beta <\pi /4$, but not if $\pi /4<\beta <\pi /2$.

\begin{figure}
\includegraphics{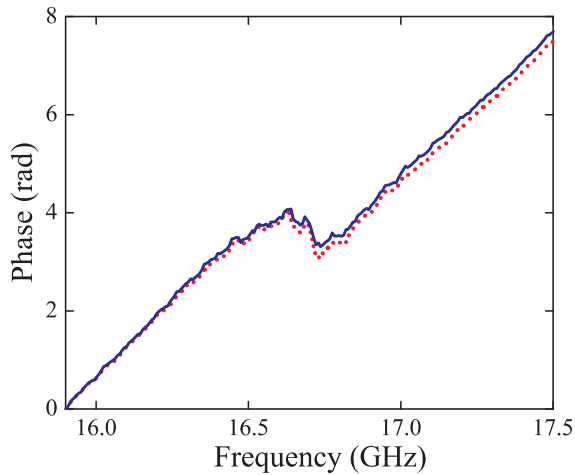}
\caption{Phase data (dotted line) and the result of the amplitude-phase
Kramers-Kronig relation applied to the transmission data (solid line).}
\label{kkdata1}
\end{figure}

In Fig. \ref{kkdata1}, we display the Kramers-Kronig transformation of Eq. %
\ref{kk-ph} applied to the amplitude data shown in Fig. \ref{beta}(a) with $%
d_{0}=1.34d/c$ ($d_{0}$ was determined by comparing the result of the
integral transformation with the actual phase data far from the
half-waveplate frequency). The actual phase data are also shown in this plot
for comparison. Clearly, the transformation works quite well at $\beta
\simeq \pi /4-0.087$, confirming that the transfer function does not have
any complicating zeros in the upper half-plane at this angle. However, this
simple transformation cannot produce the radically different phase data
obtained for $\beta \simeq \pi /4+0.087$ (shown in Fig. \ref{beta}(b)).

The interference between terms in $\tilde{H}(\omega )$ arises because of the
two (polarization) modes available to light in this system. The birefringent
photonic crystal allows a coupling between these modes, and energy can flow
from one to the other. This feature is common to all systems that display
superluminal effects. In absorptive media energy is scattered from the
incident field into other directions by fluorescence, while in transparent
active media two frequencies are present and the medium provides a coupling
between them. In passive, reflective systems energy is transferred between
the incoming and reflected waves. Since the vector formulation of the K-K
relations presented above can apply to any pair of modes, any medium that
displays this coupling between an incident mode and another mode should also
display superluminal effects. This result extends the conclusion of Bolda et
al., that the (scalar) K-K relations imply superluminal group velocities
must exist in systems that are absorptive over some range of frequencies 
\cite{Bolda1993}.

In conclusion, we have shown that superluminal and even negative group
velocities can exist as a result of interference, instead of absorption or
reflection. Moreover, it is evident from this discussion that causal
superluminal propagation can potentially be observed in any system in which
input energy can \textquotedblleft escape\textquotedblright\ into other
modes. Our results also demonstrate an example of an unusual application of
the amplitude-phase Kramers-Kronig relations; for the system we have
described, an infinitesimal adjustment of parameters radically affects the
validity of a simple transformation between the amplitude and phase.

This work was supported by ARO (grant number DAAD19-02-1-0276), ONR and NSF.
We thank the UC Berkeley Radio Astronomy Laboratory, in particular Dr. R.
Plambeck and Professor W. J. Welch for lending us the VNA. JMH thanks the
support from Instituto do Mil\^{e}nio de Informa\c{c}\~{a}o Qu\^{a}ntica,
CAPES, CNPq, FAPEAL, PRONEX-NEON, ANP-CTPETRO.

\end{document}